\documentclass[%
aps,prb,superscriptaddress,twocolumn,floatfix,10pt,longbibliography]{revtex4-2}
\usepackage[utf8]{inputenc}
\usepackage{amsmath,amssymb}
\usepackage{empheq}
\usepackage{float}
\usepackage{lipsum}
\usepackage{mathtools,cuted}
\usepackage{esvect}
\usepackage{multirow}
\usepackage{xcolor}
\usepackage{soul}
\usepackage[export]{adjustbox}
\usepackage{graphicx}
\usepackage{dcolumn}
\usepackage{bm}
\usepackage{hyperref}
\usepackage{physics}
\usepackage[mathlines]{lineno}
\usepackage[capitalise]{cleveref}
\usepackage{bbm}
\usepackage{orcidlink}
\usepackage{babel}
\usepackage{booktabs}
\usepackage{pifont}

\usepackage{pdfpages}
\usepackage{etoolbox} 

\makeatletter
\patchcmd{\@outputpage@head}{\@ifx{\LS@rot\@undefined}{}{\LS@rot}}{}{}{}
\makeatother

\newcommand{\kk}[0]{\mathbf{k}}

\makeatletter
\def\bbl@set@language#1{%
	\edef\languagename{%
		\ifnum\escapechar=\expandafter`\string#1\@empty
		\else\string#1\@empty\fi}%
	\@ifundefined{babel@language@alias@\languagename}{}{%
		\edef\languagename{\@nameuse{babel@language@alias@\languagename}}%
	}%
	\select@language{\languagename}%
	\expandafter\ifx\csname date\languagename\endcsname\relax\else
	\if@filesw
	\protected@write\@auxout{}{\string\select@language{\languagename}}%
	\bbl@for\bbl@tempa\BabelContentsFiles{%
		\addtocontents{\bbl@tempa}{\xstring\select@language{\languagename}}}%
	\bbl@usehooks{write}{}%
	\fi
	\fi}
\newcommand{\DeclareLanguageAlias}[2]{%
	\global\@namedef{babel@language@alias@#1}{#2}%
}
\makeatother

\DeclareLanguageAlias{en}{english}
\DeclareLanguageAlias{English}{english}
\DeclareLanguageAlias{Englisch}{english}
\DeclareLanguageAlias{EN}{english}
\DeclareLanguageAlias{en-US}{english}

\begin{document}
	
\title{Fractality-induced Topology}
	
\author{L. Eek}
\author{Z.~F.~Osseweijer}
\author{C. Morais Smith}
\affiliation{Institute of Theoretical Physics, Utrecht University, Utrecht, 3584 CC, Netherlands}
 
\date{\today}
	
\begin{abstract}
Fractal geometries, characterized by self-similar patterns and non-integer dimensions, provide an intriguing platform for exploring topological phases of matter. In this work, we introduce a theoretical framework that leverages isospectral reduction to effectively simplify complex fractal structures, revealing the presence of topologically protected boundary and corner states. Our approach demonstrates that fractals can support topological phases, even in the absence of traditional driving mechanisms such as magnetic fields or spin-orbit coupling. The isospectral reduction not only elucidates the underlying topological features but also makes this framework broadly applicable to a variety of fractal systems. Furthermore, our findings suggest that these topological phases may naturally occur in materials with fractal structures found in nature. This work opens new avenues for designing fractal-based topological materials, advancing both theoretical understanding and experimental exploration of topology in complex, self-similar geometries.
\end{abstract}
	
\maketitle{}

\textit{Introduction.} Symmetry-protected topological phases of matter have proved to be an important theme in condensed-matter physics. They provide a robust framework that allows for the characterization of a system based solely on the symmetries that are present. The exploration of these novel phases of matter started with the discovery of the quantum Hall effect, where the topology is driven by magnetic field, breaking time-reversal symmetry \cite{klitzing_new_1980, haldane_model_1988}. This was followed by the discovery of the quantum spin Hall effect, where the magnetic field is replaced by the spin-orbit coupling, which preserves time-reversal symmetry \cite{kane_quantum_2005, kane_z_2005}. Moreover, strain or curvature can be used to induce the quantum valley Hall effect because strain and curvature produce pseudo-magnetic fields on the different valley degrees of freedom \cite{xiao_valley-contrasting_2007, levy_strain-induced_2010}. A significant development was the realization that crystalline symmetries can also be leveraged to protect topological phases, yielding topological crystalline insulators and, consequently, crystalline higher-order topological insulators (HOTIs) \cite{fu_topological_2007, fang_new_2015, slager_space_2013, van_miert_higher-order_2018,  benalcazar_electric_2017,schindler_fractional_2019}. The latter system in $D$ dimensions does not exhibit robust modes on its $(D-1)$-dimensional boundary, but instead hosts modes on its $(D-d)$-dimensional boundaries, where $d \geq 2$ is the order of the HOTI.

Therefore, there is a plethora of different mechanisms that can drive topology. In the quantum Hall systems, an (effective) magnetic field drives topology in different sectors of the Hamiltonian. By imposing additional crystalline symmetries, the quantum Hall-type phases can be extended from two to three dimensions to obtain a second-order 3-dimensional HOTI, where combined crystal and time-reversal symmetry protect the topological phase. An example of this is topology protected by a mirror-graded Chern number, which predicts hinge states \cite{schindler_higher-order_2018}. Other types of HOTIs are systems in obstructed atomic limits \cite{bradlyn_topological_2017} and systems with a quantized multipole moment \cite{benalcazar_electric_2017}, which have $0-$dimensional corner states and are $D-$th order HOTIs in $D$ dimensions. In both of these classes, the topology is driven by a staggering (or `breathing') of the hopping parameters.

Fractals, characterized by self-similarity and non-integer dimensions, have recently emerged as a new frontier in the study of topological systems \cite{canyellas_topological_2024, conte_fractal-lattice_2024, lage_quasi-one-dimensional_2023, manna_noncrystalline_2022, manna_higher-order_2022, osseweijer_haldane_2024, li_fractality-induced_2023, kempkes_design_2019}. In particular, fractal geometries like the Sierpiński carpet and gasket have demonstrated their capacity to host robust, topologically protected modes. Nevertheless, these studies have typically involved systems that are already topologically non-trivial before being placed onto a fractal lattice. 

Here, we propose a fundamentally different mechanism by which fractality itself can induce topological phases of matter. Our work explores how the self-similar nature of fractal unit cells can lead to the emergence of topological corner states, even when there are no explicit magnetic field, spin-orbit coupling, or staggering in the hopping parameters. This suggests that fractality itself can serve as a driving mechanism for topological behavior, opening new avenues for the exploration of topological phases in novel geometries.

\begin{figure}
    \centering
    \includegraphics[width=\linewidth]{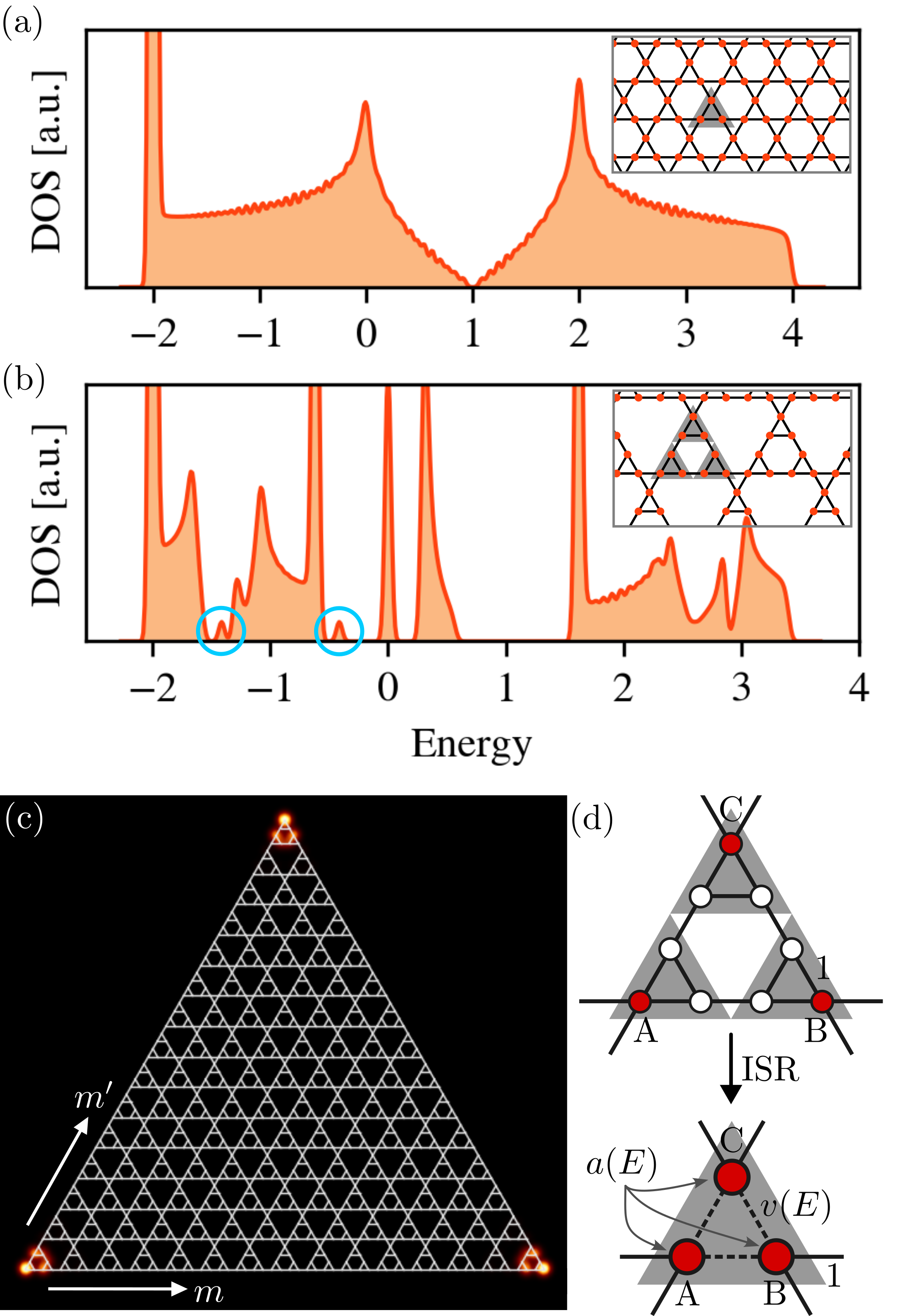}
    \caption{(a) DOS for a kagome lattice system (i.e. without fractal geometry). The inset shows the bulk lattice (b) DOS for the Sierpiński-fractal unit cell structure. The blue circles highlight in-gap modes, and the inset shows the bulk lattice structure. (c) Spatial distribution of the topologically protected corner modes, indicated by the second blue circle in (b). (d) The isospectral reduction of a first-generation Sierpínski unit cell. Red sites indicate the set $S$, white sites denote $\overline{S}$. The resulting unit cell is that of a breahting kagome lattice with intracell hopping $v(E)$ and onsite potential $a(E)$. The energy is measured in units of $t$.}
    \label{fig:DOS}
\end{figure}

\textit{Model.} We consider the simplest possible tight-binding Hamiltonians of the form
\begin{equation}
    H = t\sum_{\langle ij \rangle \in \Lambda} c^\dagger_i c^{}_j, \label{eq:H}
\end{equation}
i.e. a particle on a lattice $\Lambda$ with only nearest-neighbour hopping $t$. As a first example, we consider the kagome lattice \cite{kempkes_robust_2019}, for which we show in \cref{fig:DOS}(a) the density of states (DOS) as a function of energy $E$, which is measured in units of $t$. The inset shows the corresponding bulk lattice structure. In the absence of staggered hopping, the kagome Hamiltonian has a gapless spectrum, with a flat band at $E=-2t$, at which the DOS diverges, and a Dirac point at $E=t$. Breathing versions of the kagome lattice have been extensively studied, yielding corner modes and fractionalized corner charges \cite{kempkes_robust_2019, kunst_lattice_2018, benalcazar_quantization_2019}. Nevertheless, in the absence of breathing, the kagome lattice is (semi)-metallic and does not host topological modes. Figure~\ref{fig:DOS}(b) depicts the DOS of a lattice built from unit cells comprised of first-generation Sierpínski fractals, as shown in the inset. From now on, we will refer to this lattice as a ``Sierpínski-kagome lattice'', owing to it being constructed by replacing the triangular structure in a kagome lattice by a Sierpínski fractal structure. The DOS depicts multiple flat bands, but more importantly, reveals that the spectrum is gapped for certain filling factors. Since the Hamiltonian in \cref{eq:H} only contains one energy scale, the hopping $t$, the existence of bandgaps in the fractal lattice can not be attributed to the Hamiltonian itself. Instead, the bandgaps are a consequence of the introduction of holes in the lattice. Moreover, two gaps host in-gap states, indicated by blue circles in \cref{fig:DOS}(b). The spatial profile of the states corresponding to the right circle is displayed in \cref{fig:DOS}(c), clearly revealing localized corner states. 

\textit{Self-similarity. } The spectral features of a system inherit the self-similar nature of the fractal unit cells from which it is composed. To quantitatively describe this behavior, we introduce the concept of isospectral reduction (ISR) \cite{bunimovich_isospectral_2014}. The ISR is obtained by considering the partitioning of a matrix (or graph) into a set $S$ and $\overline{S}$, and then reducing the original problem to an effective problem on the subset $S$. Formally, the ISR of a matrix $H$ is given by
\begin{equation}
    \mathcal{R}_S(H,E) = H_{SS} - H_{S\overline{S}}\left( H_{\overline{SS}} - E\mathbb{I} \right)^{-1} H_{\overline{S}S}.
\end{equation}
It converts the linear eigenvalue problem $H\boldsymbol{\psi} = E\boldsymbol{\psi}$ into a nonlinear one $\mathcal{R}_S(H,E)\boldsymbol{\psi}_S = E\boldsymbol{\psi}_S$, where $\boldsymbol{\psi}_S$ is the $S$ component of $\boldsymbol{\psi}$. Importantly, performing an ISR does not forego spectral information of $H$. Recently, the ISR has been used in the context of latent symmetries to design flat-band lattices \cite{morfonios_flat_2021} and to explain ``accidental degeneracies'' \cite{rontgen_latent_2021}.
Moreover, latent symmetries were also used to construct latent versions of the Su-Schrieffer-Heeger (SSH) model \cite{rontgen_latent_2023} and of the non-Hermitian SSH model \cite{eek_emergent_2024}, and to build higher-order topological crystalline insulators protected by latent crystalline symmetries \cite{eek_higher-order_2024}.

For fractal lattices, the ISR can be used to relate lattices of different generations to each other. A similar approach has been applied before \cite{domany_solutions_1983, kempkes_design_2019}, but did not touch upon the presence of topological states. 
In the case of the first-generation Sierpínski-kagome lattice, we take the sites indicated in red in \cref{fig:DOS}(d) to be the set $S$. The remainder of the sites form $\overline{S}$. If one then performs the ISR, $\mathcal{R}_S(H,E)$ takes the form of a breathing kagome Hamiltonian with intercell hopping $1$, intracell hopping $v(E) = E/\left[(E^2 -1)(E-2) \right]$, and onsite potential $a(E)= 2(E^2-E-1)/\left[(E^2 -1)(E-2) \right]$. Consequently, modes at different energies are subject to an effective Hamiltonian with different parameters. The effective system resultant from performing the ISR on the first-generation Sierpínski-kagome lattice is depicted in \cref{fig:DOS}(d). For $v(E)\neq 1$, this effective kagome Hamiltonian breathes and may drive higher-order topology. In the SM we present similar equations relating $n$-th generation lattices to $(n-1)$-th generation lattices.

\begin{figure*}
    \centering
    \includegraphics[width=\linewidth]{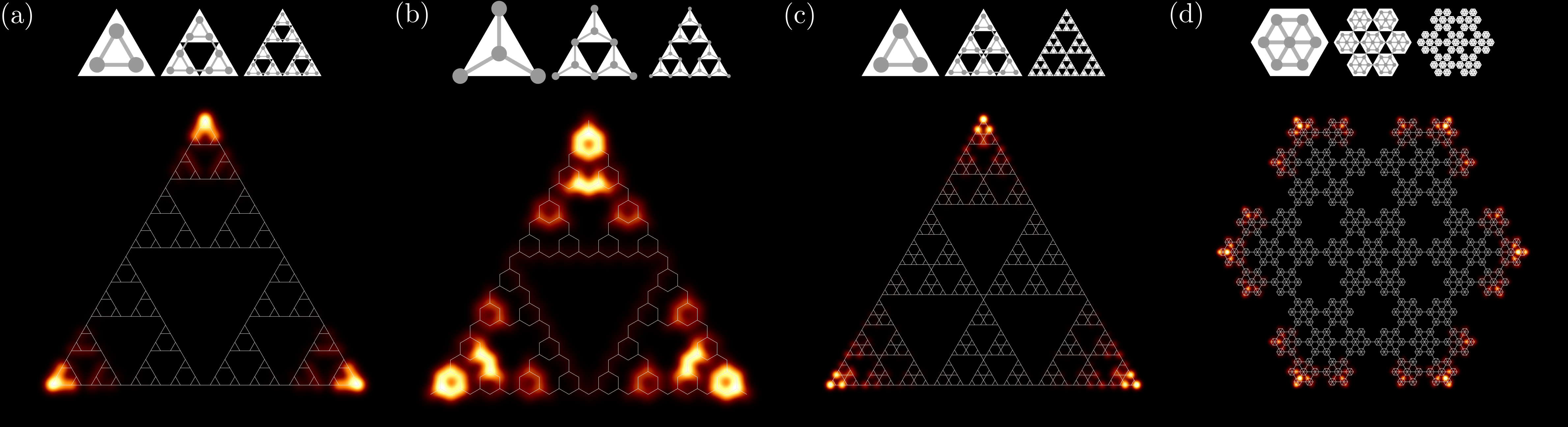}
    \caption{Spatial distribution of corner state wave-functions on fractal lattices. The heat map denotes the summed $|\psi|^2$ of a set of topological corner states. Top line of the figure shows how the fractals, from which the lattices are obtained, can be iteratively generated. (a) Fourth-generation fractal flake based on the construction method employed to obtain the fractal lattice in \cref{fig:DOS}(b-d). (b) Third-generation lattice obtained from an alternate construction method to (a), as can be seen at the top of the figure. This same method was employed in \cite{kempkes_design_2019, osseweijer_haldane_2024}. (c) Third-generation fractal lattice based on the Pascal triangle mod $3$. (d) Third-generation lattice construction based on the hexaflake fractal. The lattice is obtained by putting seven sites in each polygon that composes the fractal.}
    \label{fig:fractals}
\end{figure*}

\textit{Topological characterization. }The systems treated in this work exhibit a rotational symmetry. On the level of the Bloch Hamiltonian, $n$-fold rotations are represented by $\hat{C}_n h(\kk) \hat{C}_n^{-1} = h(D_{C_n} \kk)$, where $\hat{C}_n$ rotates the sites in the unit cell by $2\pi/n$ radians and $D_{C_n}$ rotates the momentum $\kk$ by the same amount. The higher-order topology of such systems is characterized in terms of rotational invariants, introduced in Ref.~\cite{benalcazar_quantization_2019}. The Sierpínski-kagome Hamiltonian satisfies a $C_3$-symmetry and is therefore characterized by two integers $\chi^{(3)} \equiv ([K_1^{(3)}], [K_2^{(3)}])$, which are obtained from the symmetry eigenvalues of the Bloch-functions at the high-symmetry point $\mathbf{K}$ (see SM). When $\chi^{(3)} \neq \mathbf{0}$, the system is in an obstructed atomic limit, characterized by a finite bulk dipole moment and/or fractional corner charge. For the set of states highlighted by a blue circle in \cref{fig:DOS}(b), we have $\chi^{(3)} = (1,0)$, indicating that they are topological.

The self-similarity of the fractal lattices allows for a second approach to obtain the topological phase of the system. Since repeated applications of the ISR relate the fractal lattice to an effective breathing kagome lattice, one can use properties of the latter. In order to do so, we must first obtain the corner-state energies. For the breathing kagome lattice, the corner modes sit at $E_c=0$ (onsite potential is zero for the kagome lattice). For the effective kagome model, this yields the condition for corner mode energy $E_c=a(E_c)$, where $a(E)$ is the effective onsite potential. Next, one must check whether this state is topological. The breathing kagome model is topological if the intracell hopping is smaller in magnitude than the intercell hopping. Hence, for the effective model we require $|v(E_c)|<1$. If this requirement holds for a solution to $E_c=a(E_c)$, then there will be topological corner modes at that energy $E_c$.

\textit{Anatomy of the corner states. } There is an abundance of novel states on fractal lattices. For instance, Refs.~\cite{conte_fractal-lattice_2024, biswas_designer_2023} treat the different types of compact localized states (CLS) present on fractals. Similar states also form on the lattices considered here. Instead, we focus on topological corner-like states. In the previous section, we showed that by successively applying the ISR, the fractal lattices can be mapped to an effective breathing kagome model. An expression for the corner states in the kagome model (which is exact in the thermodynamic limit) was given in Ref.~\cite{kunst_lattice_2018}. Since the eigenstates of the ISR are equal to the full eigenstates of $H$ projected on $S$ [i.e. $\mathcal{R}_S(H,E) \psi_S = E \psi_S$], based on Ref.~\cite{kunst_lattice_2018} we obtain
\begin{equation}
    \ket{\Psi_\text{corner}}_S = \mathcal{N} \sum_{m,m'} \left[- v(E_c) \right]^{m+m'} c^\dagger_{A,m,m'} \ket{0}.
\end{equation}
Here, $m$ and $m'$ are the unit cell indices in the $\mathbf{a}_1$ $=(1,0)^T$ and $\mathbf{a}_2$ $=(1/2,\sqrt{3}/2)^T$ directions, respectively. For clarity, $m$ and $m'$ have been indicated in \cref{fig:DOS}(c). Furthermore, $A$ indicates the $A$-sublattice of the kagome Hamiltonian, i.e. the bottom left site in \cref{fig:DOS}(d). The solution can be extended to $\overline{S}$ by realizing that $\ket{\psi}_S$ and $\ket{\psi}_{\overline{S}}$ are related: $\ket{\psi}_{\overline{S}} = (E-H_{\overline{SS}})^{-1} H_{\overline{S}S} \ket{\psi}_S$,
which yields the full solution $\ket{\psi} = (\ket{\psi}_S,\ket{\psi}_{\overline{S}})^T$ to $H$. The last equality is derived in detail in the SM \footnote{We recognize that obtaining the full eigenstates requires a matrix inversion, which is a computationally costly operation. However, in the case of the kagome type lattices, $H_{\overline{SS}}$ is a blockdiagonal matrix with the individual (identical) blocks being $3^{n+1} \times 3^{n+1}$ matrices (here $n$ is the generation of the fractal lattice). This significantly simplifies the computation.}. By rotational symmetry, similar solutions exist for the two other corners.

\textit{Fractals \& other models. } 
There exist multiple methods for constructing lattices from a Sierpínski gasket. One possibility is to construct a fractal lattice by taking a Sierpínski gasket, putting three sites in each triangle, and then connecting nearest neighbours, see \cref{fig:fractals}(a). Another approach is to put a site in the center and on the corners of each triangle in a gasket and subsequently connecting the nearest neighbours. This approach is shown in \cref{fig:fractals}(b), yielding a honeycomb-like lattice. This version of the Sierpínski was studied in the presence of Haldane-type next-nearest neighbour hopping in Refs.~\cite{osseweijer_haldane_2024, li_fractality-induced_2023} and on an atomic simulator platform in Ref.~\cite{kempkes_design_2019}. Nevertheless, neither of these works investigated the higher-order topology of the system. In the SM, we show that, by using a similar procedure as for the system in \cref{fig:fractals}(a), this system can be reduced to an effective breathing kagome lattice, elucidating the nature of its topological corner states. 

These fractal unit cells can be used to construct flakes that are fractal-like, conform \cref{fig:DOS}(b)-(c), or flakes that are a fractal themselves, \cref{fig:fractals}. Consider constructing a flake out of three unit cells that compose a $n$-th generation Sierpínski-kagome lattice. We obtain a flake that is a generation $(n+1)$ Sierpínski gasket. Figure.~\ref{fig:fractals}(a) shows a $4$-th generation Sierpínski gasket, with three topological corner states. Invoking topological properties is based on a bulk-boundary correspondence (BBC). From a bulk topology perspective, the fractals in \cref{fig:fractals}(a) and (b) consist of three unit cells, all three of them at the `boundary' of the system. Nevertheless, we can still heuristically argue based on a BBC. Since the corner states are exponentially localized, they do not `see' the largest and second-largest holes. Consequently, these holes could be filled and the corner states would still be solutions of the lattice. The `filled system' is similar to the one considered in Figs.~\ref{fig:DOS}(b)-(c), for which the BBC holds. The caveat here is that the localization length $\ell$ of a corner state should be smaller than the distance from the corner to a large hole. Since $\ell$ is inversely proportional to the gap in which the state sits, the corner states in large gaps survive on a fractal flake.

The presented analysis does not restrain itself solely to lattices based on the Sierpínski gasket. A direct generalization can be made by considering the \textit{Pascal triangle mod $k$} fractal, where $k$ is a prime. This fractal is iteratively constructed in a similar manner to the Sierpínski gasket, not by taking 2 copies as a side of the next generation, but taking $k$ copies \footnote{Another way to view this: The Sierpínski triangle is constructed by taking 3 identical copies of generation $n$ to construct generation $n+1$. Meanwhile for a Pascal triangle mod $k$, one has to take $k(k+1)/2$ copies of generation $n$ to construct $n+1$.}. This makes the Sierpínski gasket a special case: it is a Pascal triangle mod $2$. The Pascal triangle family of fractals has a Hausdorff dimension given by $d_H = \log_k\left[ k(k+1)/2\right]$. Figure~\ref{fig:fractals}(c) shows a lattice constructed from a $3$-rd generation Pascal triangle mod $3$, hosting topological corner modes. The three aforementioned lattices all have a $C_3$-symmetry and, through the ISR, can be related to an effective breathing kagome lattice. This does not mean that topology induced by fractality is constrained to systems that reduce to a kagome lattice. On the contrary, it is applicable to a wide range of systems. Figure~\ref{fig:fractals}(d) displays a lattice based on the hexaflake fractal, which has $d_H = \log 7/ \log 3 \approx 1.77$. Its boundary is a Koch curve, and its interior contains Koch snowflakes. The hexaflake has a $6$-fold rotation symmetry, such that its topology should be characterized by the $\chi^{(6)}$ invariant \cite{benalcazar_quantization_2019}. In the SM, we show how through the ISR the hexaflake lattice can be reduced to the Kekulé lattice, a breathing honeycomb lattice. This system is known to exhibit topology, explaining the topological corner states displayed in \cref{fig:fractals}(d) \cite{freeney_edge-dependent_2020}. Detailed calculations for all fractals in \cref{fig:fractals} are performed in the SM.

\begin{figure}[t]
    \centering
    \includegraphics[width=\linewidth]{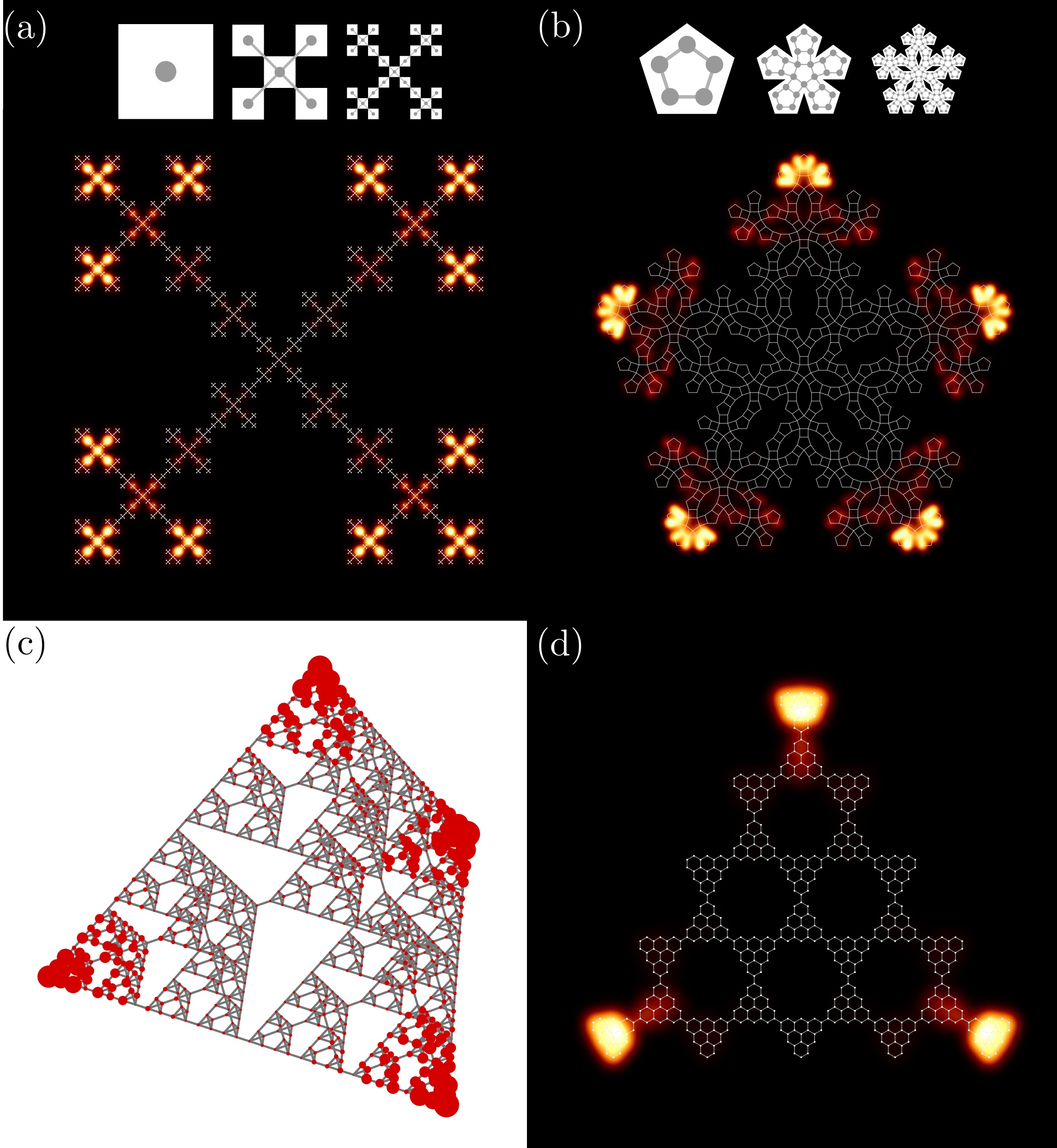}
    \caption{Spatial distribution of corner state wave-functions. The heat map denotes the summed $|\psi|^2$ of a set of topological corner states. (a) Fourth-generation Viczek ($4$-flake) fractal, the lattice is obtained by replacing the squares directly by sites. (b) Third-generation pentaflake ($5$-flake) fractal, an example of a $5$-fold symmetric fractal structure. (c) Fourth-generation Sierpínski tetrahedron fractal. The lattice is obtained by placing four sites in each tetrahedron in a similar way to the Sierpínski gasket presented in \cref{fig:fractals}(a). The size of the red balls denotes the amplitude of the wave-function on that given site. (d) Small triangular flake of $[4]$triangulene, in this case $|\psi|$ of the corner states is represented by the heatmap instead of $|\psi|^2$. This way its better visible that the wave-function exponentially decays into the interior and is not (compactly) localized on the outer cells.}
    \label{fig:outlook}
\end{figure}

\textit{Generalizations. }We demonstrated how fractality can be a driving mechanism for higher-order topological phases of matter. By considering lattices where the unit cell was constructed from a $n$-th generation fractal. The kagome lattice is a natural example, as its unit cell could be seen as a zeroth-generation Sierpínski gasket. In \cref{fig:DOS}(b) and \cref{fig:fractals}(a), we showed how taking a first- and third-generation Sierpínski gasket as the unit cell leads to the emergence of topological corner states. The analysis was then extended to different types of lattices, based on different fractals, such as the family of Pascal triangle fractals and the hexaflake. The hexaflake itself is part of a larger family of fractals, namely the $N$-flakes. These are constructed in a similary manner to the hexaflake, but by taking an $N$-sided polygon instead of a hexagon, and surrounding one central polygon by $N$ copies of itself to obtain the next generation \footnote{a variation of the $N$-flake leaves out the central polygon}. A special example of this is the $4$-flake or the Viczek fractal, which is depicted in \cref{fig:outlook}(a). Tight-binding Hamiltonians based on the Viczek fractal move away from being a lattice and tend to look more like `trees'. Nevertheless, they exhibit corner states. Furthermore, by employing the ISR, different generations of the Viczek fractal can be related, introducing an effective breathing hopping structure after the ISR. Larger lattices constructed from lower generation Viczek fractals show the same behavior.

$N$-flakes also represent a platform to study topological phases in Hamiltonians with `forbidden' symmetries. Indeed, $\mathbb{R}^2$ can only be tiled periodically by systems with a $2$-, $3$-, $4$-, or $6$-fold rotation symmetry. $N$-flakes enjoy a $N$-fold symmetry such that for $N\notin\{2,3,4,6\}$, they form a complement to quasiperiodic systems. In \cref{fig:outlook}(b), we present an example of this with the pentaflake ($5$-flake), displaying exponentially localized states on each of its corners. 

Moreover, the analysis presented in this work does not restrain itself to two spatial dimensions: the Sierpínski tetrahedron is the 3-dimensional counterpart of the gasket, from which we can construct a lattice by placing 4 sites in each tetrahedron and connecting nearest neighbours, see \cref{fig:outlook}(c). The reduction scheme here is identical to the one proposed in \cref{fig:DOS}(d), but we reduce onto a tetrahedron. Repeated application of the ISR maps the Sierpínski-tetrahedron lattice into an effective breathing pyrochlore lattice, which has been shown to host topological corner modes \cite{ezawa_higher-order_2018}.

Fractal, or fractal-like, structures are good candidates to explore novel quantum phases of matter. Even though true fractal objects only exist in the reigns of mathematical theory, finite generation fractal structures are pervasive in nature. One could think of a Lieb lattice, characteristic of cuprate perovskites, as a lattice build from first-generation Sierpínski carpets \cite{lieb_two_1989}. More recently, the band structures of metal-organic frameworks, highly tunable building blocks for complex lattices, have been analysed \cite{kumar_two-dimensional_2018}. The versatility of such systems makes them a suitable candidate to explore fractal topology. Honeycomb structures formed by HgTe or CdSe nanocrystals \cite{beugeling_topological_2015} also offer a promising platform to investigate fractal behaviour. Alternatively, triangulene, a honeycomb structure with its sites replaced by graphene nanoplatelets, presents itself as another suitable candidate \cite{ortiz_theory_2023, catarina_broken-symmetry_2023, turco_magnetic_2024}. Figure~\ref{fig:outlook}(d) displays a small flake of $[3]$triangulene, together with corner states. Saliently, $[3]$triangulene hosts corner states, while $[2]$triangulene does not. This behaviour is explained by the expressions for the effective hopping parameters upon performing the ISR and is further elaborated upon in the SM.

\textit{Conclusion. }In this work, we have demonstrated that fractal geometries provide a rich platform for the emergence of topological phases, offering a robust framework that is applicable across a broad spectrum of fractal systems. By extending the theory to various fractal structures, such as Sierpiński gaskets, Viczek, pentaflakes, and $N$-flakes, we have shown that the interplay between fractality and topology is not only theoretically intriguing, but also potentially realizable in real-world materials, such as metal-organic frameworks, triangulenes and honeycomb nanocrystals. These findings suggest that fractal-based topological materials could naturally occur in a variety of physical systems, reflecting the self-similar patterns found in nature. The generality of our approach opens up new avenues for experimental exploration, particularly in the study of naturally occurring and artificially designed fractal materials, which may harbor exotic topological states and functionalities. This broad applicability underscores the significance of fractal topology in advancing our understanding of condensed-matter systems and potentially guiding the development of future quantum materials.\\

\textit{Acknowledgements. }The authors thank Anouar Moustaj and Rodrigo Arouca for useful discussions. All authors acknowledge the research program “Materials for the Quantum Age” (QuMat) for financial support. This program (registration number 024.005.006) is part of the Gravitation program financed by the Dutch Ministry of Education, Culture and Science (OCW).

\bibliography{refsAPS}

\newpage

\onecolumngrid


\section*{Supplementary material}

# Supplemental Material: Fractality-induced Topology

L. Eek,[1] Z. F. Osseweijer,[1] and C. Morais Smith[1]

[1]*Institute of Theoretical Physics, Utrecht University, Utrecht, 3584 CC, Netherlands*

(Dated: November 19, 2024)

## I. ISOSPECTRAL REDUCTION

The ISR is obtained by partioning a Hamiltonian $H$ into a set $S$ and $\overline{S}$, i.e.

$$H\psi = \begin{pmatrix} H_{SS} & H_{S\overline{S}} \\ H_{\overline{S}S} & H_{\overline{S}\overline{S}} \end{pmatrix} \begin{pmatrix} \psi_S \\ \psi_{\overline{S}} \end{pmatrix} = E \begin{pmatrix} \psi_S \\ \psi_{\overline{S}} \end{pmatrix}. \tag{S1}$$

This yields two equations

$$\begin{aligned} H_{SS}\psi_S + H_{S\overline{S}}\psi_{\overline{S}} &= E\psi_S, \\ H_{\overline{S}S}\psi_S + H_{\overline{S}\overline{S}}\psi_{\overline{S}} &= E\psi_{\overline{S}}. \end{aligned} \tag{S2}$$

Substituting the second in the first yields the ISR as defined in the main text,

$$\mathcal{R}_S(H,E)\psi_S \equiv \left( H_{SS} + H_{S\overline{S}} \left[ H_{\overline{S}\overline{S}} - E \right]^{-1} H_{\overline{S}S} \right) \psi_S = E\psi_S.$$

In the main text we obtain an expression for the corner states $\psi_S$j. This can be extended to $\psi_{\overline{S}}$ by rewriting the second line of Eq. (S2),

$$\psi_{\overline{S}} = (E - H_{\overline{S}\overline{S}})^{-1} H_{\overline{S}S} \psi_S. \tag{S3}$$

## II. HIGHER-ORDER TOPOLOGY

All systems treated in this work obey a rotational symmetry. The (higher-order) topology of such systems is characterized in terms of rotational invariants, introduced in Ref. [1]. In two dimensions, rotations of a lattice by $2\pi/n$ rad around some point are represented on the level of the Bloch Hamiltonian by

$$\hat{C}_n h(\mathbf{k})\hat{C}_n^{-1} = h(D_{C_n}\mathbf{k}), \tag{S4}$$

where $D_{C_n}$ rotates the crystal momentum $\mathbf{k}$ by $2\pi/n$ rad. At high symmetry points, $\mathbf{k} = \mathbf{\Pi^{(n)}}$, in the Brillouin zone we have

$$\left[ \hat{C}_n, h\left( \mathbf{\Pi^{(n)}} \right) \right]_{-} = 0, \tag{S5}$$

such that $\hat{C}_n$ and $h(\mathbf{k})$ commute at a high-symmetry point and, therefore, share an eigenbasis, such that

$$\hat{C}_n \left| u\left( \mathbf{\Pi^{(n)}} \right) \right\rangle = \Pi_p^{(n)} \left| u\left( \mathbf{\Pi^{(n)}} \right) \right\rangle. \tag{S6}$$

Here, $\Pi_p^{(n)} = \exp\{2\pi i(p-1)/n\}$ with $p \in \{1, 2, \ldots n\}$, since the eigenvalues of $\hat{C}_n$ are the $n$-th roots of unity. From these eigenvalues, we construct

$$\left[ \Pi_p^{(n)} \right] \equiv \#\Pi_p^{(n)} - \#\Gamma_p^{(n)}, \tag{S7}$$

where $\#\Pi_p^{(n)}$ denotes the number of filled bands with eigenvalue $\Pi_p^{(n)}$ and $\mathbf{\Gamma} = \mathbf{0}$ is the gamma point. A linear independent set of $[\Pi_p^{(n)}]$ then constitutes the rotational invariant $\chi^{(n)}$ for a $n-$fold rotation-symmetric Hamiltonian [1]. For $n = 2, 3, 4, 6$, we obtain

$$\begin{aligned} \chi^{(2)} &= \left( \left[ X_1^{(2)} \right], \left[ Y_1^{(2)} \right], \left[ M_1^{(2)} \right] \right), \\ \chi^{(4)} &= \left( \left[ X_1^{(2)} \right], \left[ M_1^{(4)} \right], \left[ M_2^{(4)} \right] \right), \\ \chi^{(3)} &= \left( \left[ K_1^{(3)} \right], \left[ K_2^{(3)} \right] \right), \\ \chi^{(6)} &= \left( \left[ M_1^{(2)} \right], \left[ K_1^{(3)} \right] \right). \end{aligned} \tag{S8}$$

The set of rotational invariants characterizes the higher-order topology of the gaps in $n$-fold rotational symmetric lattices. Physical quantities such as bulk dipole moments and corner charges can be derived from these invariants. Expressions are given in Ref. [1].

## III. DETAILED CALCULATIONS FOR SIERPÍNSKI-KAGOME LATTICES

### A. Recursive parameters

In this section we give an example on how to derive the recursively defined effective parameters for fractal models. To this extent, we consider Sierpínski-kagome type models. In order to derive these equations, we start from the more general Hamiltonian

$$H^{(n)}(a,v,w) = a \sum_i c_i^\dagger c_i + v \sum_{\langle ij \rangle \in \triangle} c_i^\dagger c_j + w \sum_{\langle ij \rangle \notin \triangle} c_i^\dagger c_j, \tag{S9}$$

where the lattice is a $n$-th generation Sierpinski-kagome fractal, like the ones depicted in Fig. S1. The nearest-neighbour (NN) hopping inside a shaded triangle ($\langle i,j \rangle \in \triangle$) is set to $v$ while NN hopping between triangles ($\langle i,j \rangle \notin \triangle$) is set to $w$. In the main text, we examine the case where $a = 0$ and $v = w = t$, i.e. $H^{(n)}(0,t,t)$.

Consider $H^{(2)}(a,v,w)$, which is depicted in Fig. S1(a) for $a = 0$ and $v = w = 1$. The zoom-in inset shows a single unit cell. Defining $S$ as the sites indicated in red and taking the ISR, one obtains the system in Fig. S1(b). Here, the solid-line (black) hoppings remain equal $w$ ($= 1$), while the dashed hoppings become

$$\begin{aligned}
v_1(E) &= \frac{v^2 w(E - a + v - w)}{[(E-a)^2 - v^2 + vw - w^2](E - a - v - w)} \\
&\equiv f(E, v, w, a),
\end{aligned} \tag{S10}$$

and the new on-site potential is

$$\begin{aligned}
a_1(E) &= a + \frac{2v^2}{3}\Bigg[ \frac{2(E-a) + 2v - w}{(E-a)^2 - v^2 + vw - w^2} \\
&\qquad\qquad + \frac{1}{E - a - v - w}\Bigg] \\
&\equiv g(E, v, w, a).
\end{aligned} \tag{S11}$$

Consequently, the governing Hamiltonian is now $H^{(1)}[a_1(E), v_1(E), w_1(E)]$.

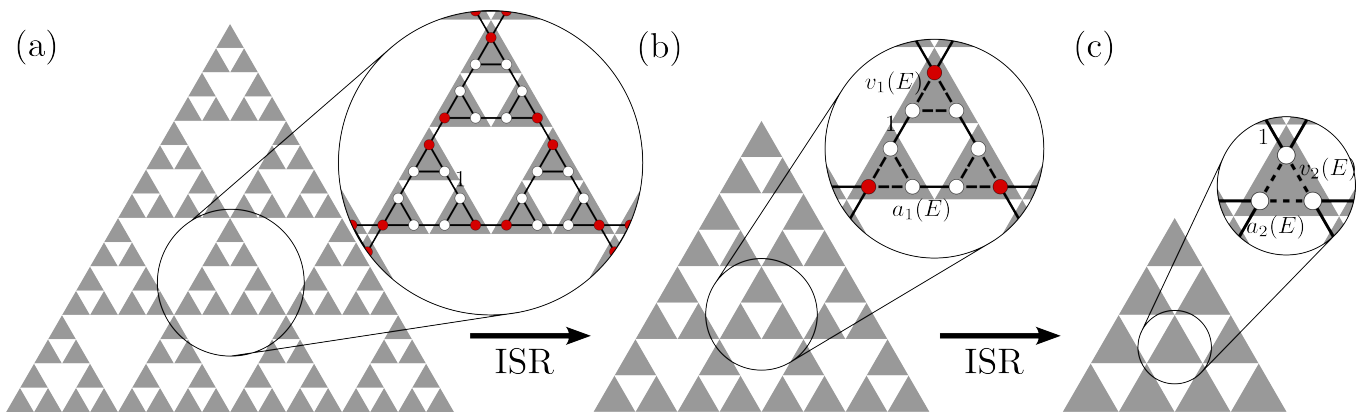

FIG. S1. Relation of different generations of $H^{(n)}(a,v,w)$ through ISR. (a) $H^{(2)}(0,1,1)$. Here all hoppings are equal. The set $S$ is indicated by the red sites. Upon taking an ISR on the red sites, (b) is obtained. (b) $H^{(1)}(a_1(E), v_1(E), 1)$. The system is now a breathing fractal lattice of one generation lower than (a). By applying yet another ISR on the red sites, one obtains (c): a breathing kagome lattice $H^{(0)}(a_2(E), v_2(E), 1)$.

Reducing once further with $S$ now being the red sites in the zoom-in inset of Fig. S1(b), one obtains Fig. S1(c). Here, the solid black hoppings are still equal to $w$, but the dashed hoppings are now equal to

$$\begin{aligned} v_2(E) &= f\left[E, v_1(E), w, a_1(E)\right] \\ &= f\left[E, f(E, v, w, a), w, g(E, v, w, a)\right], \end{aligned} \tag{S12}$$

and the on-site potential is given by

$$\begin{aligned} a_2(E) &= g\left[E, v_1(E), w, a_1(E)\right] \\ &= g\left[E, f(E, v, w, a), w, g(E, v, w, a)\right]. \end{aligned} \tag{S13}$$

The reduced parameters are thus recursively defined:

$$\begin{aligned} v_{n+1}(E) &= f(E, v_n, w_n, a_n) \\ w_{n+1}(E) &= w_n(E) = w_0 \\ a_{n+1}(E) &= g(E, v_n, w_n, a_n), \end{aligned} \tag{S14}$$

with $v_0 = v$, $w_0 = w$ and $a_0 = a$. Therefore, we conclude that by successively employing $n$ times the ISR, we can reduce the $n$-th generation fractal unit cell lattices with equal hoppings to a breathing kagome lattice with energy-dependent parameters. The results in the main text can be obtained by requiring $v_0 = w_0 = t$ and $a_0 = 0$.

## B. Band topology and corner states

In this section, we analyze the bulk topology of the Sierpiński-kagome type lattices. In Fig. S2, we depict the band structures for a triangular lattice, a kagome lattice (zeroth-generation Sierpiński-kagome), a first- and second-generation Sierpiński-kagome lattice. In these lattices, all hoppings are set to be equal to $t$, such that the Hamiltonians for Figs. S2(b)-(d) are given by $H^{(n)}(0, t, t)$, with $n = 1, 2, 3$, as defined through Eq. (S9).

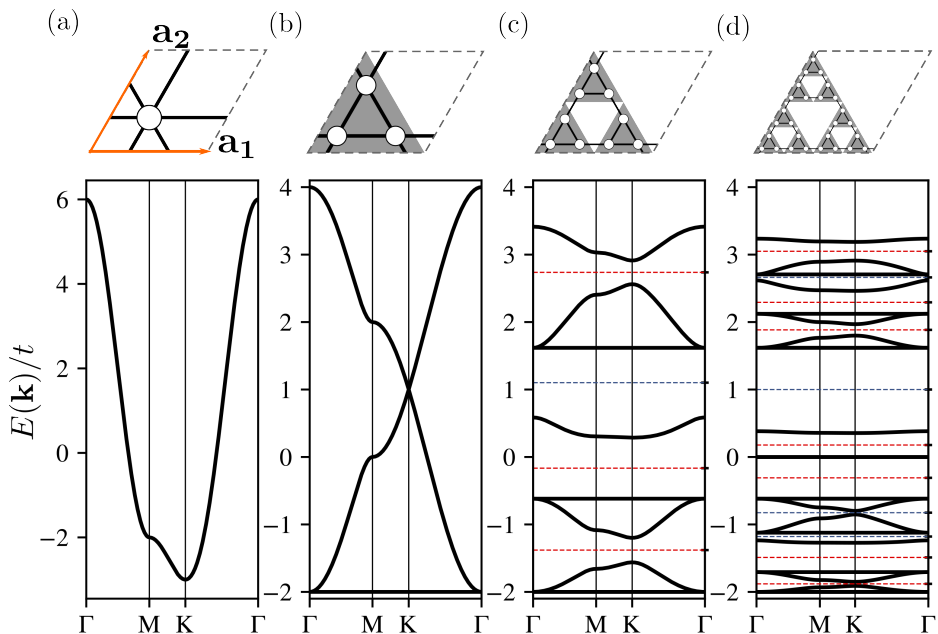

FIG. S2. Bulk spectra for (a) a triangular lattice with hopping $t$ and (b-d) $H^{(0,1,2)}(0, t, t)$. The spectra in (c,d) are gapped and for each gap the rotational invariant $\chi^{(3)}$ is indicated by the color of the dashed line. Red indicates $\chi^{(3)} = (1, 0)$, while blue represents $\chi^{(3)} = \mathbf{0}$. Unit cells of the different lattices are depicted on top of the spectra.

The lattices in Figs. S2(a)-(d) have a 3-fold rotation symmetry, i.e. their Bloch Hamiltonians satisfy $\hat{C}_3 h(\mathbf{k}) \hat{C}_3^{-1} = h(D_{C_3} \mathbf{k})$. Since the spectra in Figs. S2(a),(b) are gapless, they do not show any topological modes. On the contrary, higher-generation lattices show multiple bandgaps, both topological and trivial. Specifically, $H_{(1)}(0, t, t)$ has 9 bands

and 4 bandgaps, and $H_{(2)}(0, t, t)$ has 27 bands and 11 bandgaps (some of the flatbands are degenerate). Each of these bandgaps has been indicated in Fig. S2(c) and S2(d) by a horizontal line; red (blue) indicates a topologically non-trivial (trivial) gap, characterized by $\chi^{(3)} = (1, 0)$ [$\chi^{(3)} = (0, 0)$]. The same results are also summarized in Table I.

A non-trivial $\chi^{(3)}$ indicates the presence of topological corner modes. However, there is a caveat: if the dipole moment $\mathbf{P}^{(n)}$ is non-zero, there will be trivial edge states in the gap. For 3-fold symmetric systems, the dipole moment is given by [1]

$$\mathbf{P}^{(3)} = \frac{2e}{3} \left( \left[ K_1^{(3)} \right] + 2 \left[ K_2^{(3)} \right] \right) (\mathbf{a_1} + \mathbf{a_2}).$$ (S15)

For $\chi^{(3)} = (1, 0)$, this corresponds to $\mathbf{P}^{(3)} = (2e/3)(\mathbf{a_1} + \mathbf{a_2})$. Since the dipole moment does not vanish, there will be trivial dipole-induced edge states in the gap, together with the topological corner modes. The gapped nature of the spectra corresponding to $H^{(n=1,2)}(0, t, t)$ can be understood from the previous section: these systems can be mapped into a breathing kagome lattice. The breathing kagome lattice, which is described by $H^{(0)}(0, v, w)$, is topological for $|v| < |w|$, characterized by $\chi^{(3)} = (1, 0)$. This is the same value of $\chi^{(3)}$ that is observed for the Sierpiński-kagome lattices, confirming their similarity.

| $H^{(1)}(0, t, t)$ | | $H^{(2)}(0, t, t)$ | | | |
|---|---|---|---|---|---|
| $n$ | $\chi^{(3)}$ | $n$ | $\chi^{(3)}$ | $n$ | $\chi^{(3)}$ |
| 3 | $(1, 0)$ | 6 | $(1, 0)$ | 18 | $(0, 0)$ |
| 5 | $(1, 0)$ | 8 | $(1, 0)$ | 21 | $(1, 0)$ |
| 6 | $(0, 0)$ | 9 | $(0, 0)$ | 23 | $(1, 0)$ |
| 8 | $(1, 0)$ | 11 | $(0, 0)$ | 24 | $(0, 0)$ |
| | | 14/17 | $(1, 0)$ | 26 | $(1, 0)$ |

TABLE I. Value of the rotational invariant $\chi^{(3)}$ for $H^{(1)}$ and $H^{(2)}$ at fillings $n$ for which the spectrum is gapped. These invariants have also been indicated in blue [$\chi^{(3)} = (0, 0)$, trivial] and red [$\chi^{(3)} = (1, 0)$, topological] in Figs. S2(c),(d).

To corroborate the results obtained from the bulk spectra, we analyze systems with open boundary conditions (OBC). To this extent, we consider triangular flakes of first- and second-generation Sierpínski-kagome unit cells. The corresponding OBC spectra are depicted in Figs. S3(a) and S3(b), respectively. Here, the green bands represent the bulk band-widths, corresponding to the bands in Figs. S2(c) and S2(d). Furthermore, bulk-like states have been colored black, edge-like states are gray, and the red boxes represent corner-like states. Figure S3(c,e) show the three corner states (one for each corner) of the two different gaps of the first-generation lattice highlighted in Fig. S3(a). Similarly, Figs. S3(d), (f)-(h) show the corner states for four of the six gaps of the second-generation lattice indicated in Fig. S3(b). Since the state depicted in Fig. S3(h) has a much larger localization length than the other states, it is shown on a larger flake. Furthermore, the corner states indicated in red and green exist on both the first- and second-generation lattice, at the same energy. This can be understood as follows. The corner states on the first-generation lattice have a sufficiently small localization length, that upon 'placing' them on the second-generation lattice, they decay before hitting the larger holes. For this reason, they are also (approximate) solutions on higher generation lattices. This argument holds as long as the decay length of a corner state is shorter than the length scale introduced by the new holes of the higher-generation fractal.

Saliently, in Fig. S3(a) [S3(b)], only two (six) sets of topological states have been indicated for the first- (second-) generation lattice. Meanwhile, the bulk invariant calculations, see Figs. S2(c),(d) and Table I, predict three (seven) topological gaps, each hosting topological states. To better understand this discrepancy, we first remark that similar features were also observed in breathing kagome lattices [2]. In breathing kagome lattices [with Hamiltonian $H^{(0)}(a, v, w)$], a combination of edge-like and corner-like states is observed in the gap that is opened by modulating the hopping parameters. The system is topological for $|v| < |w|$. Furthermore, for $v = w$, the gap closes at $E = a + v$, while for $v = -w$, the gap closes at $E = a$. These two conditions can be combined in the statement that the breathing kagome model has a gap-closing at

$$E = a + (v/2) \left[ 1 + \text{sign}(v/w) \right]$$ (S16)

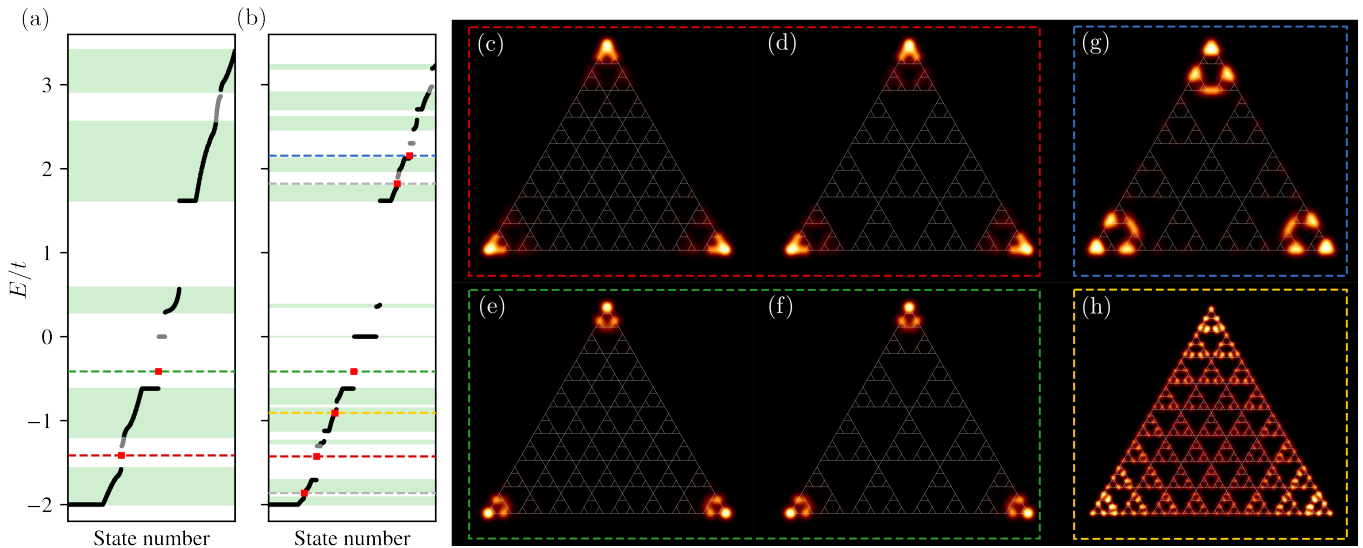

FIG. S3. Open boundary condition spectra for (a) $H^{(2)}(0, t, t)$ and (b) $H^{(3)}(0, t, t)$. Green bands represent the bulk bandwidths. Bulk-like states have been colored black, edge-like states are gray, and the red boxes represent corner-like states. (c-h) Corner states of $H^{(2,3)}(0, t, t)$. The colored border around the flakes indicates the energy of the state in (a) and (b).

This condition can be extended to obtain the gap-closing conditions for the fractal lattices by replacing the parameters with their energy-dependent counterparts. For a $n$-th generation unit cell, the gap-closing condition is given by

$$E^* = a_n(E^*) + \frac{v_n(E^*)}{2} \left\{ 1 + \text{sign}[v_n(E^*)/w] \right\}, \tag{S17}$$

where $a_n(E)$, $v_n(E)$, and $w_n(E)$ are defined in Section III A. The solutions $E^*$ of the above equation constitute the energies at which the Sierpiński-kagome lattices exhibit a gap-closing. Solving Eq. (S17) for the second-generation Sierpiński-kagome lattice yields the energies $E^*$ listed in the first column of Table II. These energies have been identified with the amount of filled bands $n$ they correspond to in the same column. Since the breathing kagome model is topological for $|v| < |w|$, one naively expect the Sierpiński-kagome model to be topological for $|v_2^*| < |w|$ [$v_2^* \equiv v_2(E^*)$]. Indeed, evaluating this condition yields a direct agreement with $\chi^{(3)}$ for that same gap, as can be seen in the second column of Table II. Nevertheless this does not solve the discrepancy regarding the corner states.

The discrepancy can be understood by realizing that corner states in the kagome model do not sit at the gap closing energy [Eq. (S16)]. Instead, they always sit at $E = a$, regardless of the sign of $v/w$. For the fractal models, this extends to the condition

$$\tilde{E} = a_n(\tilde{E}). \tag{S18}$$

Note that $E^* = \tilde{E}$ when $\text{sign}[v_n(E^*)/w] = -1$, since the latter implies that the second term in Eq. (S17) vanishes.

Consider now again the second-generation lattice [Fig. S2(d)]. Solving Eq. (S18) yields 10 solutions, the same amount of solutions as $E^*$. The lowest solution $E^* = -1.852$ corresponds to the lowest-lying gap, i.e. a filling of $n = 6$. The corresponding solution of Eq. (S18) equals $\tilde{E} = -1.862$. The associated effective hopping parameters $v_2^* \equiv v_2(E^*) = 0.900$ and $\tilde{v}_2 \equiv v_2(\tilde{E}) = 0.459$ are both smaller in magnitude than $w = 1$. Hence, we expect this gap to be topological and host corner states. This expectation is confirmed by Fig. S3(b), depicting corner states (red squares) in the lowest gap. On the contrary, effective hoppings $v_2^*$ and $\tilde{v}_2$ obtained from the solutions to Eqs. (S17) and (S18) at filling $n = 11$ are $v_2^* = 1.211$ and $\tilde{v}_2 = 0.831$. From $v_2^*$, we would expect a trivial gap since its magnitude is larger than $|w| = 1$, which is in agreement with the value of $\chi^{(3)} = (0, 0)$. Nevertheless, $|\tilde{v}_2|$ is smaller than 1, suggesting a topological gap, which agrees with the presence of corner modes.

| $n$ | $E^*$ | $\tilde{E}$ | $\|v_2^*\| < \|w\|$ | $\chi^{(3)} \neq \mathbf{0}$ | $\|\tilde{v}_2\| < \|w\|$ | Corner states |
|---|---|---|---|---|---|---|
| 6 | $-1.852$ | $-1.863$ | ✓ | ✓ | ✓ | ✓ |
| 8 | $-1.426$ | $-1.426$ | ✓ | ✓ | ✓ | ✓ |
| 9 | $-1.177$ | $-1.177$ | ✗ | ✗ | ✗ | ✗ |
| 11 | $-0.842$ | $-0.907$ | ✗ | ✗ | ✓ | ✓ |
| 14/17 | $-0.416$ | $-0.416$ | ✓ | ✓ | ✓ | ✓ |
| 18 | $+1.216$ | $+1.125$ | ✓ | ✗ | ✗ | ✗ |
| 21 | $+1.892$ | $+1.822$ | ✓ | ✓ | ✓ | ✓ |
| 23 | $+2.155$ | $+2.155$ | ✓ | ✓ | ✓ | ✓ |
| 24 | $+2.701$ | $+2.701$ | ✗ | ✗ | ✗ | ✗ |
| 26 | $+2.918$ | $+2.895$ | ✓ | ✓ | ✗ | ✗ |

TABLE II. Numerical values of $E^*$ and $\tilde{E}$ for $H^{(2)}(0,1,1)$ and comparisons between $v_2^*$, $\tilde{v}_2$ and the presence of bulk topology or corner modes.

The comparison of $v_2^*$ and $\tilde{v}_2$ for the different gaps of $H^{(2)}(0,t,t)$ has been itemized in Table II. Here, we observe that the inequality $|v_2^*| < |w|$ coincides with the predictions of $\chi^{(3)}$, while $|\tilde{v}_2| < |w|$ is indicative of the presence of corner modes. Intuitively, this may be understood by considering that the corner modes (if they exist) sit at $E = \tilde{E}$, and therefore are described by an effective Hamiltonian with parameters evaluated at $\tilde{E}$. The ordinary kagome model's parameters are the same for all states/energies. Therefore, the gap closing and existence of corner modes can be captured by the same inequality $|v| < |w|$ [3]. For the fractal lattice models, this is no longer the case.

## IV. EXPLICIT ISOSPECTRAL REDUCTIONS

### A. Honeycomb Sierpínski

The Sierpínski honeycomb lattice, introduced in Fig. 2(b) of the main text, can be related to the Sierpínski kagome lattice [Fig. 2(a) of the main text]. This is done by performing an isospectral reduction, with $S$ denoting the central sites of the triangles. These sites are indicated in red in Fig. S4. The energy-dependent parameters in the reduced model are given by

$$v(E) = \frac{1}{E}, \qquad a(E) = \frac{3}{E}. \tag{S19}$$

These parameters can now be used as a starting point for the analysis presented in Section III.

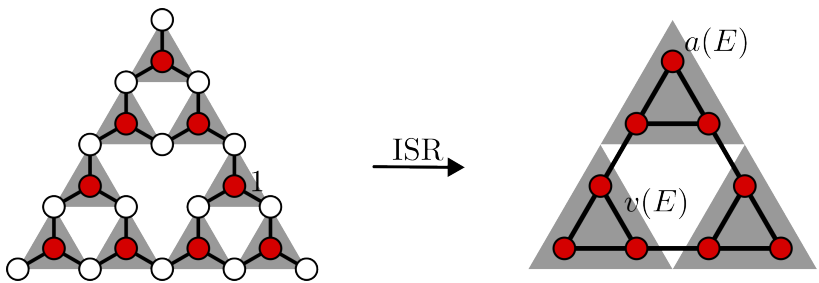

FIG. S4. Isospectral reduction of the Sierpínski honeycomb lattice to the Sierpínski kagome lattice.

### B. Pascal triangle mod 3

Similar to the reduction method for the Sierpínski-kagome lattice, outlined in Fig. 1(d) of the main text, the Pascal triangles can also be reduced to breathing kagome lattices. Figure S5 shows how to choose the sites $S$ and $\overline{S}$. The expressions for $a(E)$ and $v(E)$ can be readily obtained by explicitly performing the ISR. This method can be iteratively applied to higher generation Pascal fractals by taking the corners of the largest scale triangles as the sites $S$.

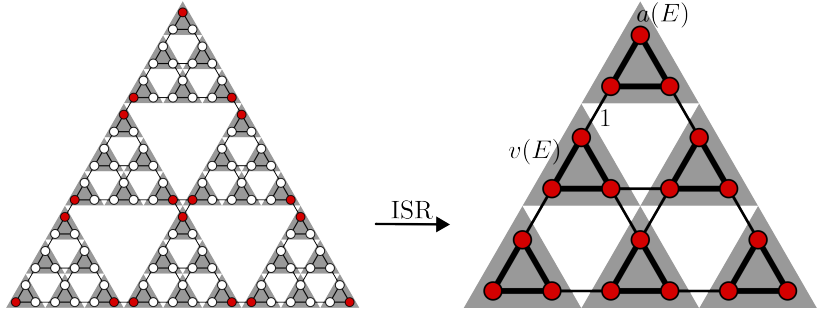

FIG. S5. Isospectral reduction of the Pascal triangle mod 3 to a breathing kagome lattice.

### C. Hexaflake

The lattices based on the hexaflake fractal can be mapped into a Kekulé lattice [4] by repeated application of the ISR. This is done by considering 7 small hexagons that compose a larger hexagon, and reducing on the corner sites. This process is depicted in Fig. S6, where the red sites form the set $S$. In this specific case, besides introducing a breathing feature, the ISR also introduces longer range intracell hoppings $v_2(E)$ and $v_3(E)$. Nevertheless, this lattice still hosts topological corner modes. See, for example, Ref. [5]. The explicit expressions for the effective onsite potential $a(E)$ and hoppings $v_i(E)$, $i \in \{1, 2, 3\}$, are quite lengthy and therefore omitted. They can be obtained by performing the ISR.

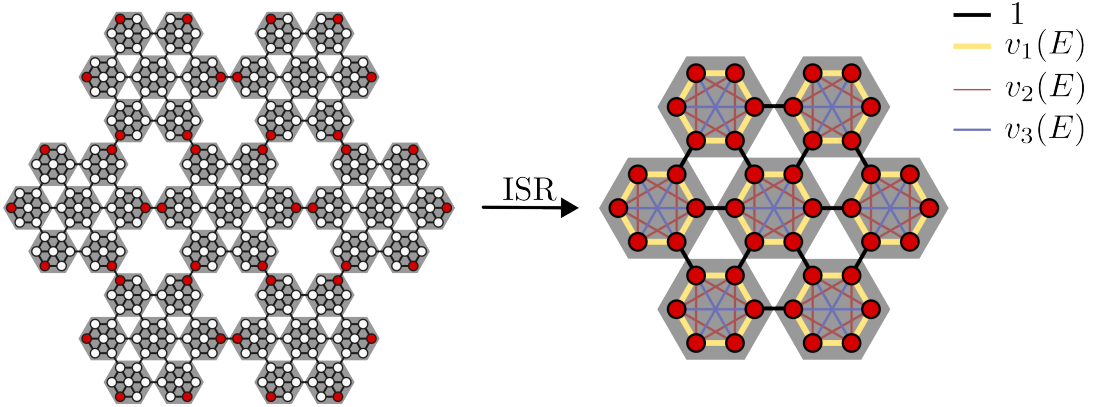

FIG. S6. Isospectral reduction of a second-generation hexaflake lattice to the Kekulé lattice. Before the ISR (left), all hoppings are equal and the onsite potential is zero. After performing an ISR (right), all sites have an effective on-site potential $a(E)$ and are connected by different (colour-coded) hoppings.

### D. Triangulene

Triangulene lattices are obtained by taking triangle-shaped graphene platelets and connecting them, as shown in Fig. S7. Depending on the size of the platelet, one obtains $[n]$triangulene. In Fig. S7 we depict the building blocks for $[1]$-, $[2]$-, and $[3]$triangulene. By performing an ISR on the corner sites of the shaded triangles, indicated in red in Fig. S7, one obtains an (energy-dependent) breathing truncated hexagonal [or $(3, 12^2)$] lattice. The effective parameters for the first three $[n]$triangulenes are given by

$$a^{[1]}(E) = \frac{2}{E}, \qquad v^{[1]}(E) = \frac{1}{E} \tag{S20}$$

$$a^{[2]}(E) = \frac{2(E^4 - 6E^2 + 7)}{E(E^4 - 7E^2 + 10)}, \qquad v^{[2]}(E) = \frac{E^2 - 1}{E(E^4 - 7E^2 + 10)} \tag{S21}$$

$$a^{[3]}(E) = \frac{2(E^8 - 13E^6 + 55E^4 - 90E^2 + 49)}{E(E^8 - 14E^6 + 65E^4 - 118E^2 + 72)}, \qquad v^{[3]}(E) = \frac{E^4 - 2E^2 - 1}{E(E^8 - 14E^6 + 65E^4 - 118E^2 + 72)} \tag{S22}$$

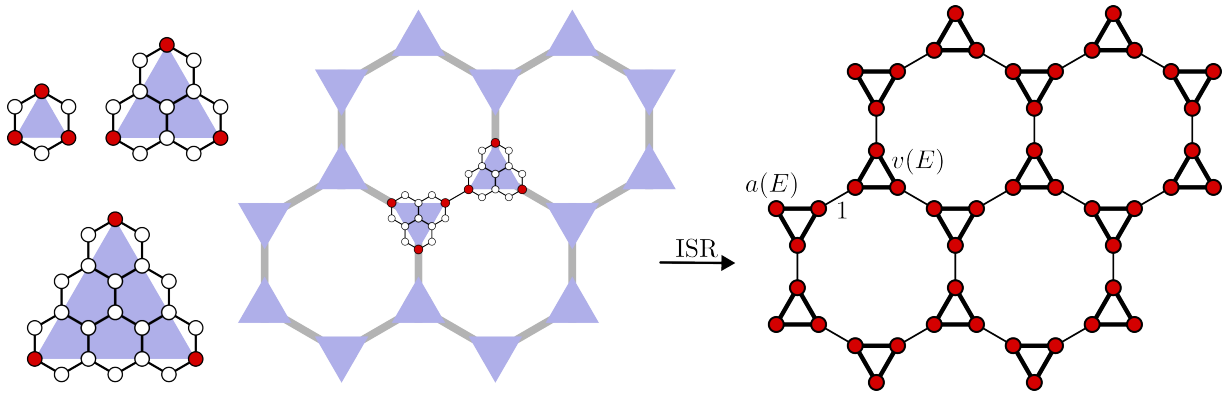

FIG. S7. Triangulene and its ISR to a breathing truncated hexagonal lattice.

To understand the existence of corner states on triangulene, we first shift our attention to corner states on the (non-latent) $(3, 12^2)$-lattice.

### 1. Corner states in the $(3, 12^2)$-lattice

We now focus on the topology of the $(3, 12^2)$-lattice which is the lattice on the right-hand side of Fig. S7 but with $v(E) = v$, i.e. no energy-dependent $v(E)$, see Fig. S8. In this case, the system is topological for $|v| < 2/3$ with corner states at $E_{\text{corner}}(v)$. The value of $E_{\text{corner}}(v)$ depends non-trivially on $v$ and can be obtained by making an Ansatz for the existence of exponentially localized corner states.

Writing out the Schrödinger equation for the wavefunctions on sites $1 - 4$, one obtains [See Fig. S8]

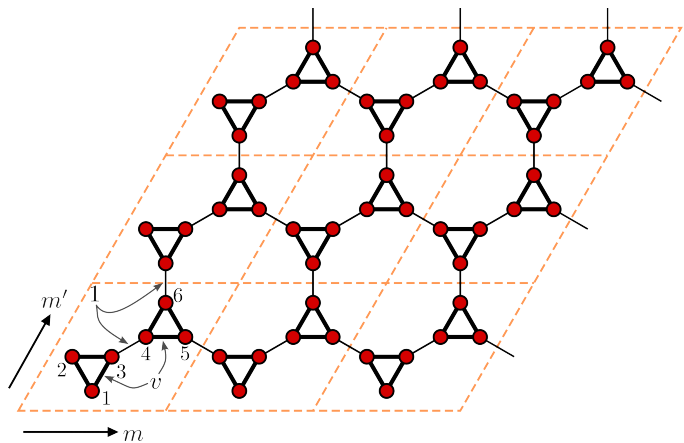

FIG. S8. A corner of a flake of a $(3, 12^2)$-lattice. Thick black lines represent a hopping of magnitude $v$, while thin black lines represent a hopping of magnitude 1. Unit cells have been indicated by orange dashed lines.

$$E\psi_1^{(1,1)} = v\left(\psi_2^{(1,1)} + \psi_3^{(1,1)}\right), \tag{S23}$$

$$E\psi_2^{(1,1)} = v\left(\psi_1^{(1,1)} + \psi_3^{(1,1)}\right), \tag{S24}$$

$$E\psi_3^{(1,1)} = v\left(\psi_1^{(1,1)} + \psi_2^{(1,1)}\right) + \psi_4^{(1,1)}, \tag{S25}$$

$$E\psi_4^{(1,1)} = \psi_3^{(1,1)} + v\left(\psi_5^{(1,1)} + \psi_6^{(1,1)}\right), \tag{S26}$$

where $\psi_i^{(m,m')}$ is the wave-function of the $i$-th site in cell $(m, m')$. Now, assume that the corner-state wavefunction is equal to one at sites 1 and 2, but vanishes at sites 5, 6 in cell $(m, m') = (1, 1)$. This yields

$$\boldsymbol{\psi}^{(1,1)} = \begin{pmatrix} 1 \\ 1 \\ \frac{2vE_{\text{corner}}}{E_{\text{corner}}^2 - 1} \\ \frac{2v}{E_{\text{corner}}^2 - 1} \\ 0 \\ 0 \end{pmatrix}, \tag{S27}$$

where the corner-state energy $E_{\text{corner}}$ is a solution to the equation

$$E^3 - vE^2 - (1 + 2v^2)E + v = 0. \tag{S28}$$

The analytical form of $E_{\text{corner}}$ is complicated, but its behavior is shown in Fig. S9(a). Writing out the Schrödinger equation for $\psi_6^{(1,1)}$ yields

$$E\psi_6^{(1,1)} = v\psi_4^{(1,1)} + \psi_1^{(1,2)}, \tag{S29}$$

from which we obtain (after substituting $\psi_5^{(m,m')} = \psi_6^{(m,m')} = 0$)

$$\psi_1^{(1,2)} = -\frac{2v^2}{E_{\text{corner}} - 1}\psi_1^{(1,1)}. \tag{S30}$$

Generally, the corner states of the $(3, 12^2)$-lattice can be written as

$$|\Psi_{\text{corner}}\rangle = \sum_{m,m'} \left[-\frac{2v^2}{E_{\text{corner}}^2 - 1}\right]^{m+m'} \begin{pmatrix} 1 \\ 1 \\ \frac{2vE_{\text{corner}}}{E_{\text{corner}}^2 - 1} \\ \frac{2v}{E_{\text{corner}}^2 - 1} \\ 0 \\ 0 \end{pmatrix} \cdot \mathbf{c}^{(m,m')\dagger}|0\rangle, \tag{S31}$$

where $\mathbf{c}^{(m,m')\dagger} \equiv \left(c_1^{(m,m')\dagger}, \ldots, c_6^{(m,m')\dagger}\right)^T$. By analogy, similar states exist on the other corners of a triangular $(3, 12^2)$-lattice flake.

### 2. Corner states in triangulene

For the $(3, 12^2)$-lattice, the corner states sit at $E = E_{\text{corner}}$. For the effective energy-dependent parameters of reduced triangulene, this expression generalizes to

$$E - a^{[n]}(E) = E_{\text{corner}}[v^{[n]}(E)], \tag{S32}$$

where $a^{[n]}(E)$ and $v^{[n]}(E)$ are defined in Eqs. (S20) to (S22). $E_{\text{corner}}(v)$ was defined in the previous section as the solution Eq. (S28) and its behavior is shown in Fig. S9(a). For ease of notation, we introduce $\Lambda(E) \equiv E_{\text{corner}}[v^{[n]}(E)] - E + a^{[n]}(E)$. Finding solutions of Eq. (S32) is identical to finding the roots of $\Lambda(E)$. We define $E_{\text{corner}}(v)$ to only exist for $|v| < 2/3$, since for $|v| > 2/3$ no topological corner states exist. Consequently, $E_{\text{corner}}$ is not a continuous

function, and therefore $\Lambda(E)$ will not be continuous either. Figure S9(b) shows $\Lambda(E)$ (and its roots) for [1], [2], and [3]triangulene. In the main text, we argue that [2]triangulene does not exhibit corner modes. Nevertheless Fig. S9(b) depicts two roots. However, these roots are at $E = \pm 1$. At this energy, [2]triangulene has a flat-band, hiding the corner modes. For [3]triangulene, the corner modes do not sit inside a flat band, but instead in a gap, and therefore can be resolved.

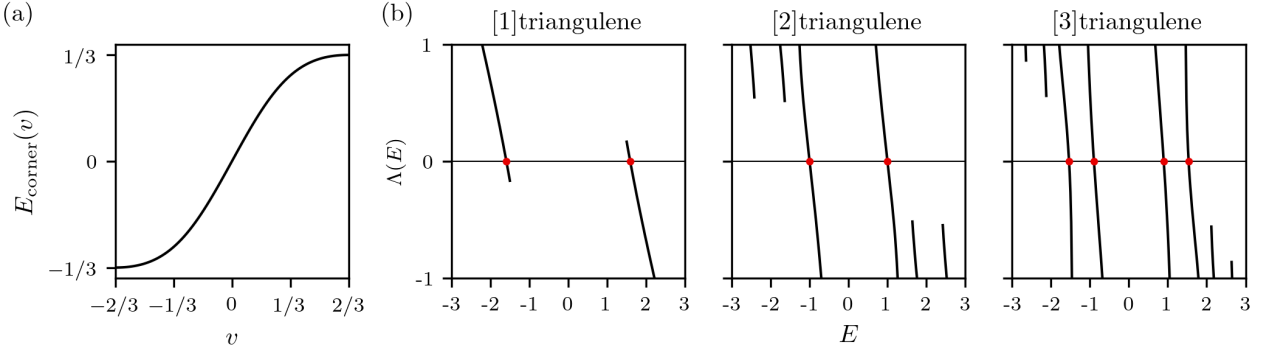

FIG. S9. (a) The corner-state energy of a $(3, 12^2)$-lattice. (b) $\Lambda(E)$ as a function of $E$ for different types of triangulene. The roots, corresponding to triangulene corner-state energies, are indicated by red dots.

The full corner states for triangulene lattices may be obtained by using Eq. (S31) to construct wave-functions on $S$, and consequently extending these wave-functions to $\overline{S}$ using Eq. (S3).

## V. INNER CORNER STATES

The larger holes of fractal lattices pose themselves as pinning points for *inner corner states* that are not localized at the outer boundaries of the lattice but instead at other corners. Similar inner corner states have also been observed in [6, 7]. Figure S10 shows some examples of corner states localizing on inner corners of different generations of the Sierpínski-kagome lattices.

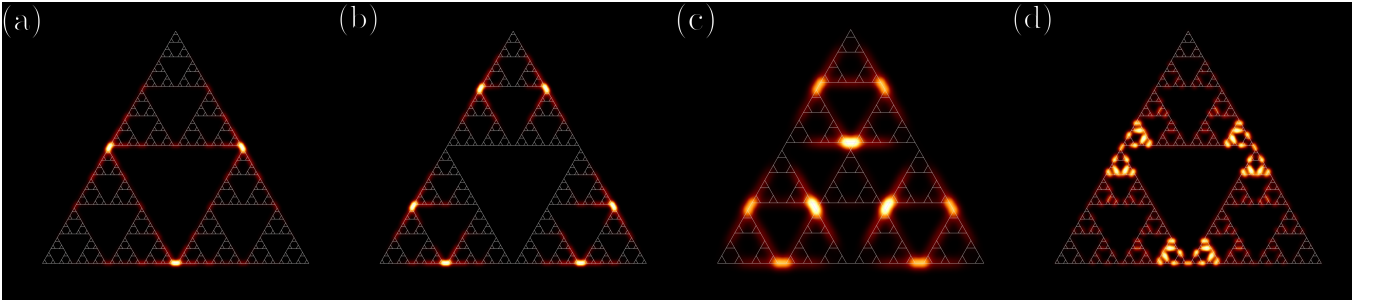

FIG. S10. Corner states at various inner corners for different generations of Sierpínski-kagome lattices.

Such corner states exist for the different lattices depicted in the main text, such as the Pascal fractals and the $N$-flakes. However, the inner corner states reside in the flat bands, and consequently mix with the compact localized states that are also present in the lattice [8]. Nevertheless, it is possible to find expressions for the wave-functions of these states. The simplest case would be an inner corner formed by two kagome lattices, as depicted in Fig. S11. Here, the size of the circles indicates the wave-function amplitude, while the color represents the sign of the wave-function, with red positive and blue negative. The full wave-function of such state is given by

$$|\psi_{\text{inner}}\rangle = \mathcal{N} \sum_x \sum_{y \leq x} \frac{1}{|x|} \left( c_{x,y,A}^\dagger - \frac{|x| + 1 - y}{|x| + 1} c_{c,y,B}^\dagger - \frac{y}{|x| + 1} c_{x,y,C}^\dagger \right) |0\rangle . \tag{S33}$$

Saliently, these states decay algebraically as $1/d$, with $d$ the distance to the inner corner. Similar expressions for inner

corner state wave-functions can be obtained for higher-generation lattices or the different fractals presented in the main text. All these wavefunctions will exhibit $1/d$ decay, as can also be seen in Fig. S10.

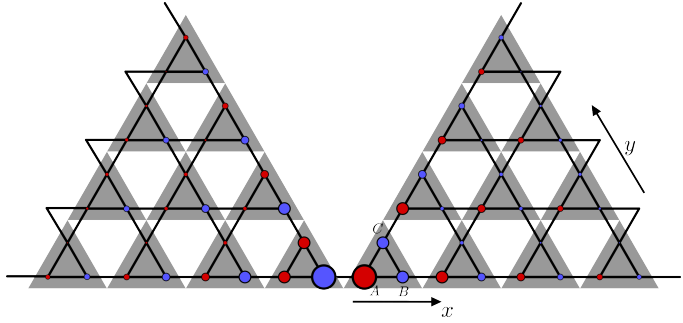

FIG. S11. Inner corner state for a kagome lattice at $E = -1$. All hoppings are equal and set to 1. The amplitude of the circles indicates the amplitude of the inner corner state wave-function, while the color represents the sign of the wave-function.

## VI. CONSTRUCTION OF FRACTAL LATTICES

There are multiple ways to construct lattices from fractal structures. In this section, we take the Sierpínski gasket as an example to construct fractal lattices. The Sierpínski is constructed iteratively by starting with a triangle, taking three copies, and forming a new triangle with a hole in it. This yields a first-generation Sierpínski gasket. Repeating this procedure by taking the first-generation fractal as a starting point yields a second-generation fraction. This process is repeated to obtain $n$-th generation Sierpínski gaskets. To construct a lattice, we select sites in the initial triangle and then iteratively construct the fractal. One example is by choosing sites on each of the three corners of the triangle. By doing so, one obtains the lattice depicted in Fig. S12(a) (here, and in the following lattices, nearest neighbors have been connected). Another method is to put a site in the center of each triangle. This yields the lattice in Fig. S12(b), which is a Sierpínski-kagome lattice. In fact, these two constructions yield lattices that are each others dual. Combining these two approaches, i.e. putting sites on the corners and in the center of a filled triangle, yields the Sierpínski-honeycomb lattice from Section IV A, see Fig. S12(c). Nevertheless, any structure could be put on the initial triangle. In Fig. S12(d), we arrange sites in a triangle within the initial triangle, yielding another Sierpínski-kagome lattice.

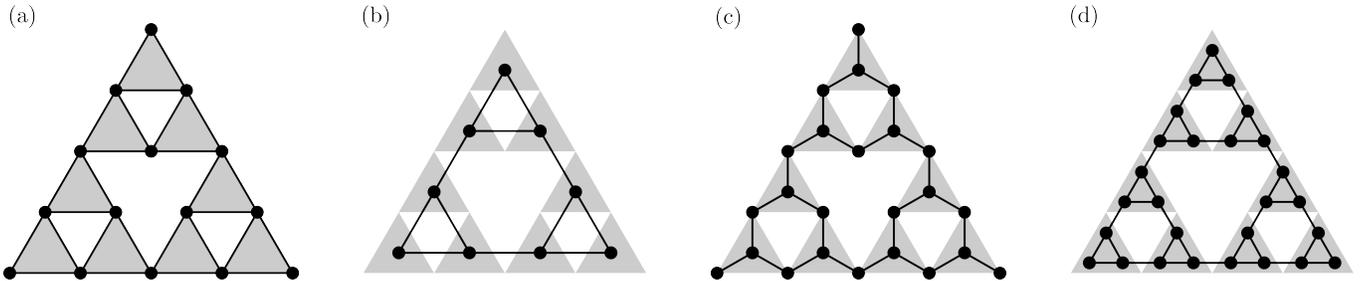

FIG. S12. Fractal lattice construction methods. (a) Putting sites on the corners of each triangle. (b) Putting sites in the center of each filled triangle. (c) Putting sites on the corners and in the center of each filled triangle. (d) Putting three triangularly arranged sites in each triangle.

Some construction methods are similar. For example, the method outlined in Fig. S12(d) could be regarded equal to the one in Fig. S12(b) except for a third-generation Sierpínski gasket instead of second. This is illustrated in Fig. S13.

(a)          (b)

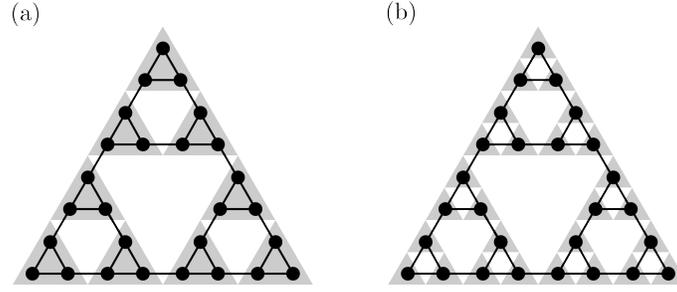

FIG. S13. Similarity of lattices obtained by (a) Putting triangularly arranged sites in each filled triangle, conform Fig. S12(d), for a second-generation Sierpínski gasket. (b) Putting a single site in the center of each triangle, conform Fig. S12(b), for a third-generation Sierpínski gasket.

---


\end{document}